\documentclass[compsoc,conference,a4paper,10pt,times]{IEEEtran}
\IEEEoverridecommandlockouts

\usepackage[utf8]{inputenc}

\newif\ifvldb
\vldbfalse

\usepackage{booktabs} 

\usepackage{listings}
\usepackage{color}
\usepackage{colortbl}
\usepackage[x11names,dvipsnames,table]{xcolor}
\usepackage{comment}
\usepackage{xspace}
\usepackage{enumerate}
\usepackage{color}
\usepackage{graphicx}
\usepackage{outlines}
\usepackage{enumitem}
\usepackage{url}
\usepackage{caption}
\usepackage{subcaption}
\usepackage{amsmath}
\usepackage{amssymb}
\usepackage{xcolor}
\usepackage{tikz,forest}
\usetikzlibrary{arrows.meta}
\usepackage{enumitem}
\setlist{nosep}
\usepackage{multirow,bigdelim}
\usepackage{float}
\usepackage[normalem]{ulem}
\usepackage{amsthm}
\usepackage{lineno}

\definecolor{midnightblue}{rgb}{0.1, 0.1, 0.44}
\definecolor{darkslateblue}{rgb}{0.28, 0.24, 0.55}
\definecolor{smokyblack}{rgb}{0.06, 0.05, 0.03}
\definecolor{xanadu}{rgb}{0.25, 0.23, 0.27}
\usepackage[framemethod=default]{mdframed}
\mdfdefinestyle{exampledefault}{%
frametitlefont=\small\bfseries,
frametitleaboveskip=1mm,
frametitlebelowskip=1mm,
rightline=true,innerleftmargin=10,innerrightmargin=10,
frametitlerule=true,
frametitlebackgroundcolor=gray!25,
skipabove=0pt,
innerleftmargin=2mm,
innertopmargin=1pt,
innerrightmargin=2mm,
innerbottommargin=4pt,
nobreak=true
}

\usepackage{amsmath}
\usepackage{amsthm}
\usepackage{semantic}

\usepackage{ragged2e}
\usepackage{bold-extra}

\newcommand{\todo}[1]{{\color{red}~\textbf{TODO: #1}}}

\newcommand{\om}[1]{{\color{blue}~~~\emph{\textbf{#1}}}}
\renewcommand{\om}[1]{}
\newcommand{\arjun}[1]{\textcolor{brown}{(Arjun: #1)}}
\renewcommand{\arjun}[1]{}

\newcommand{\tocut}[1]{\textcolor{brown}{\sout{#1}}}
\renewcommand{\tocut}[1]{}

\newcommand{\system}{{\scshape{Chorus}}\xspace}
\newcommand{\systemhead}{{\bfseries{\scshape{\large Chorus}}}\xspace}

\renewcommand{\paragraph}[1]{\vspace{1mm}\noindent\textbf{#1}}

\newcommand{\numqueries}{18,774\xspace}
\newcommand{\numtripssampled}{200 million\xspace}

\DeclareMathOperator{\Lap}{Lap}

\newcommand{\rmspcsm}{\vspace*{-0.8mm}}
\newcommand{\mydownarrow}{\rmspcsm{\hspace{.2\textwidth}$\Downarrow$}\rmspcsm}

\makeatletter
\g@addto@macro\normalsize{%
  \setlength\abovedisplayskip{-5pt}
  \setlength\belowdisplayskip{4pt}
  \setlength\abovedisplayshortskip{-5pt}
  \setlength\belowdisplayshortskip{4pt}
}

\ifvldb

\else

\fi

\newif\ifextended
\extendedtrue

\ifextended
\newcommand{\exten}[1]{#1}
\else
\newcommand{\exten}[1]{}
\fi

\ifextended
\newcommand{\extend}[2]{#1}
\else
\newcommand{\extend}[2]{#2}
\fi

\forestset{mechanisms/.style={for tree={
            base=bottom,
            child anchor=north,
            align=center,
            s sep+=1cm,
    straight edge/.style={
        edge path={\noexpand\path[\forestoption{edge},thick,-{Latex}] 
        (!u.parent anchor) -- (.child anchor);}
    },
    if n children={0}
        {tier=word, draw, thick, rectangle}
        {draw, diamond, thick, aspect=2},
    if n=1{%
        edge path={\noexpand\path[\forestoption{edge},thick,-{Latex}] 
        (!u.parent anchor) -| (.child anchor) node[pos=.2, above] {Yes\ \ \ \ \ };}
        }{
        edge path={\noexpand\path[\forestoption{edge},thick,-{Latex}] 
        (!u.parent anchor) -| (.child anchor) node[pos=.2, above] {\ \ \ \ \ No};}
        }
        }
    }
}

\AtBeginEnvironment{forest}{\baretabulars}

\usepackage{cite}
\usepackage{amsmath,amssymb,amsfonts}
\usepackage{algorithmic}
\usepackage{graphicx}
\usepackage{textcomp}
\usepackage{bmpsize}
\usepackage{xcolor}
\usepackage{lipsum}
\def\BibTeX{{\rm B\kern-.05em{\sc i\kern-.025em b}\kern-.08em
    T\kern-.1667em\lower.7ex\hbox{E}\kern-.125emX}}

\begin{document}

\date{}



\title{\system: a Programming Framework for Building \\Scalable Differential Privacy Mechanisms}

\author{\IEEEauthorblockN{Noah Johnson}
  \IEEEauthorblockA{\textit{UC Berkeley} \\
    noahj@berkeley.edu}
  \and
\IEEEauthorblockN{Joseph P. Near}
  \IEEEauthorblockA{\textit{University of Vermont} \\
    jnear@uvm.edu}
  \and
\IEEEauthorblockN{Joseph M. Hellerstein}
  \IEEEauthorblockA{\textit{UC Berkeley} \\
    hellerstein@berkeley.edu}
  \and
\IEEEauthorblockN{Dawn Song}
  \IEEEauthorblockA{\textit{UC Berkeley} \\
    dawnsong@berkeley.edu}
}






\maketitle
\thispagestyle{plain}
\pagestyle{plain}

\begin{abstract}
Differential privacy is fast becoming the gold standard in enabling statistical analysis of data while protecting the privacy of individuals. However, practical use of differential privacy still lags behind research progress because research prototypes cannot satisfy the scalability requirements of production deployments. To address this challenge, we present \system, a framework for building scalable differential privacy mechanisms which is based on cooperation between the mechanism itself and a high-performance production database management system (DBMS). We demonstrate the use of \system to build the first highly scalable implementations of complex mechanisms like Weighted PINQ, MWEM, and the matrix mechanism. We report on our experience deploying \system at \extend{Uber}{a large technology company}, and evaluate its scalability on real-world queries.



\end{abstract}

\begin{IEEEkeywords}
privacy, differential privacy, SQL queries, query rewriting, security
\end{IEEEkeywords}

\newcommand{\revisioncolor}{}

\newcommand{\rmfigspace}{}

\section{Introduction}
\label{sec:intro}

Organizations are collecting more and more sensitive information about individuals. As this data is highly valuable for a broad range of business interests, organizations are motivated to provide analysts with flexible access to the data. At the same time, the public is increasingly concerned about privacy protection. There is a growing and urgent need for technology solutions that balance these interests by supporting general-purpose analytics while guaranteeing privacy protection.

Differential privacy~\cite{dworkdifferential2006,dwork2014algorithmic} is widely recognized by experts as the most rigorous theoretical solution to this problem. Differential privacy provides a formal guarantee of privacy for individuals while allowing general statistical analysis of the data. In short, it states that the presence or absence of any single individual's data should not have a large effect on the results of a query. This allows precise answers to questions about populations in the data while guaranteeing the results reveal little about any individual. Unlike alternative approaches such as anonymization and k-anonymity, differential privacy protects against a wide range of attacks, including attacks using auxiliary information~\cite{sweeney1997weaving,DBLP:journals/corr/abs-cs-0610105,taxis,de2013unique}. 

Current research on differential privacy focuses on development of new algorithms, called \emph{mechanisms}, to achieve differential privacy for a particular class of queries. Researchers have developed dozens of mechanisms covering a broad range of use cases, from general-purpose statistical queries~\cite{dwork2006calibrating, nissim2007smooth, mcsherry2007mechanism, mcsherry2009privacy, proserpio2014calibrating, mohan2012gupt, Blocki:2013:DPD:2422436.2422449} to special-purpose analytics tasks such as graph analysis~\cite{hay2009accurate,sala2011sharing,karwa2011private,kasiviswanathan2013analyzing,Chen:2013:RMT:2463676.2465304}, linear queries~\cite{hardt2012simple, li2014data, li2010optimizing, li2015matrix, zhang2014towards, xiao2012dpcube, cormode2012differentially, qardaji2013differentially, acs2012differentially, xu2013differentially, mckenna2018optimizing}, and analysis of data streams~\cite{dwork2010differential,shi2011privacy}.


Despite extensive academic research and an abundant supply of mechanisms, differential privacy has not been widely adopted in practice. Existing applications of differential privacy in practice are limited to specialized use cases~\cite{erlingsson2014rappor, apple}.

A major challenge for the practical adoption of differential privacy is the ability to deploy differential privacy mechanisms \emph{at scale}. Today's data analytics infrastructures include industrial-grade database management systems (DBMSs) carefully tuned for performance and reliability, designed to process datasets consisting of billions of rows.

The simplest mechanisms for differential privacy, like the \emph{Laplace mechanism}~\cite{dwork2006calibrating}, answer an analyst's query by adding noise to the final result of the query. This mechanism can be easily deployed atop an existing high-performance DBMS by leveraging the DBMS to execute the analyst's query, then adding the right amount of noise to the result. Since it uses the DBMS to perform the actual data processing tasks, this approach scales well. Some existing work, such as {\sc Flex}~\cite{allegro}, takes this approach, which we call the \emph{post-processing} architecture. For appropriate mechanisms, the post-processing architecture solves the scalability problem.

However, more advanced differential privacy mechanisms require fundamental changes to the way queries execute. Summation queries, for example, require \emph{clipping} the data before summing it to control the influence of outliers. The post-processing approach is fundamentally incompatible with mechanisms like this one---it is \emph{impossible} to run the analyst's query unmodified and then achieve differential privacy by post-processing the results. Unfortunately, the vast majority of recently-developed mechanisms fall into this category (e.g.~\cite{hay2009accurate,sala2011sharing,karwa2011private,kasiviswanathan2013analyzing,Chen:2013:RMT:2463676.2465304, hardt2012simple, li2014data, li2010optimizing, li2015matrix, zhang2014towards, xiao2012dpcube, cormode2012differentially, qardaji2013differentially, acs2012differentially, xu2013differentially, mckenna2018optimizing}), and cannot be implemented using the post-processing architecture.

As a result, no scalable implementation exists for many of the exciting differential privacy mechanisms developed in recent years. The implementations which have been developed (e.g.~\cite{mohan2012gupt, proserpio2014calibrating, mcsherry2009privacy, wilson2019differentially, kotsogiannis2019privatesql}) modify or replace the DBMS with a custom engine, which is unlikely to offer performance on par with modern production DBMSs.

\paragraph{The \system Framework.}
This paper describes \system, a framework for developing and deploying cutting-edge differential privacy mechanisms at scale. \system makes it easy to develop mechanism implementations which work \emph{in cooperation with} an existing high-performance DBMS, even for mechanisms which require modifying queries or generating entirely new ones. \system supports scalable implementations by leveraging the DBMS for data processing tasks, instead of custom code. We call this the \emph{cooperative} architecture for differential privacy mechanisms.

\system provides a programming framework to support implementing mechanisms in the cooperative architecture. The framework has three major components: \textbf{rewriting}, for modifying queries to perform functions like clipping; \textbf{analysis}, for analyzing queries to determine properties like how much noise is required for differential privacy; and \textbf{post-processing}, for processing the results of executing queries. To implement a summation mechanism with clipping, for example, we can use \system's rewriting component to modify the analyst's query so that \emph{the DBMS performs the clipping} as well as the summation. With this modification, the rest of the mechanism can be implemented via analysis and post-processing of the rewritten query.

\system supports integration with \emph{any} standard SQL database. The framework is designed to facilitate working directly with SQL queries, since SQL is the most commonly used language for high-performance production DBMSs. By using a standard SQL database engine instead of a custom runtime, \system can leverage the reliability, scalability and performance of modern databases, which are built on decades of research and engineering experience.

The cooperative architecture applies to \emph{all} of the recently-developed differential privacy mechanisms---even ones that require significant changes to the way queries execute or generate entirely new queries. We demonstrate the flexibility of the approach, and of the \system framework, by implementing both simple mechanisms like summation with clipping and complex mechanisms like wPINQ, MWEM, and the matrix mechanism. In all of these implementations, \system supports scalability by moving data processing tasks to the DBMS.

\paragraph{Deployment.}
We have made \system available as open source software~\extend{\cite{chorusdownload}}{\cite{chorusdownloadanon}}, and it is designed for integration in production environments. We describe how to deploy \system to provide differential privacy in the face of untrusted analysts who may submit malicious queries, and present practical strategies for privacy budget management as part of a \system deployment.

We report on our experience deploying \system at \extend{Uber}{a large technology company\footnote{We omit the name of the company to ensure anonymity for the submission.}} for its internal analytics tasks. \system represents a significant part of the company's General Data Protection Regulation (GDPR)~\cite{gdpr} compliance efforts, and provides both differential privacy and access control enforcement. 

\paragraph{Evaluation.}
We evaluate the scalability of \system on real queries written by analysts at \extend{Uber}{the company in which \system is deployed}, using a 300 million rows sampled from the production data. Our evaluation results demonstrate that mechanism implementations built with \system are capable of scaling to real-world analysis tasks.


\paragraph{Contributions.} In summary, we make the following contributions:

\begin{enumerate}[topsep=1mm,leftmargin=4mm]
\itemsep1.0mm
\item We present the \system framework, which enables a novel \emph{cooperative architecture} for implementing differential privacy mechanisms atop high-performance DBMSs (\S~\ref{sec:ipq}).

\item We demonstrate \system's flexibility by developing scalable implementations for a number of advanced differential privacy mechanisms (\S~\ref{sec:intr-priv-quer-1}).

\item We release \system as open source~\extend{\cite{chorusdownload}}{\cite{chorusdownloadanon}} and describe how to deploy it to provide differential privacy in production settings (\S~\ref{sec:implementation}).

\item We report on our experience deploying \system to enforce differential privacy at \extend{Uber}{a large technology company}, where it processes more than 10,000 queries per day (\S~\ref{sec:uber_deployment}).


\item We demonstrate the scalability of \system by evaluating it on \numqueries real-world queries with a database of 300 million rows (\S~\ref{sec:evaluation}). 




\end{enumerate}

\tocut{\noindent The paper is organized as follows. Section~\ref{sec:backgr-diff-priv} provides background on differential privacy and a survey of existing differential privacy mechanisms. Section~\ref{sec:ipq} describes our novel architecture and the \system system. Section~\ref{sec:selection} describes automatic mechanism selection. Section~\ref{sec:intr-priv-quer-1} introduces our transformations from non-private to intrinsically private queries, and Section~\ref{sec:building-blocks} formalizes them. Section~\ref{sec:evaluation} contains our empirical evaluation using \system.}
\section{Background\tocut{: Differential Privacy}}
\label{sec:backgr-diff-priv}

Differential privacy provides a formal guarantee of
\emph{indistinguishability}.
This guarantee is defined in terms of a \emph{privacy budget} $\epsilon$---the smaller the budget, the stronger the guarantee.
The formal definition of differential privacy
is written in terms of the \emph{distance} $d(x,y)$ between two
databases, i.e. the number of entries on which they differ: $d(x,y) = |\{i
: x_i \neq y_i\}|$. Two databases $x$ and $y$ are \emph{neighbors} if $d(x,y) = 1$. A randomized mechanism $\mathcal{K} : D^n \rightarrow
\mathbb{R}$ preserves $(\epsilon, \delta)$-differential privacy if
for any pair of neighboring databases $x, y \in D^n$ and set $S$ of possible outputs:

\[\mbox{Pr}[\mathcal{K}(x)\in S] \leq e^\epsilon
  \mbox{Pr}[\mathcal{K}(y) \in S] + \delta \]

\noindent Differential privacy can be enforced by adding noise to the non-private results of a query. The scale of this noise depends on the \emph{sensitivity} of the query. The \emph{global sensitivity} of a query $f : D^n \rightarrow \mathbb{R}$ is defined as:

\[GS_f = \max_{x,y:d(x,y)=1} \lvert f(x)-f(y) \rvert \]

\noindent Importantly, differential privacy mechanisms satisfy a \emph{sequential composition} property: if $F_1$ satisfies $(\epsilon_1, \delta_1)$-differential privacy, and $F_2$ satisfies $(\epsilon_2, \delta_2)$-differential privacy, then running both $F_1$ and $F_2$ satisfies $(\epsilon_1 + \epsilon_2, \delta_1 + \delta_2)$-differential privacy.
For more on differential privacy, see Dwork and Roth~\cite{dwork2014algorithmic}.

\paragraph{Statistical queries.}
Differential privacy aims to protect the privacy of individuals in the context of \emph{statistical queries}. In SQL terms, these are queries using standard aggregation operators (\texttt{COUNT}, \texttt{AVG}, etc.) as well as histograms created via the \texttt{GROUP BY} operator in which aggregations are applied to records within each group. Differential privacy is not suitable for queries that return raw data (e.g. rows in the database) since such queries are inherently privacy-violating.

\paragraph{Mechanism design.}
Research on differential privacy has produced a large and growing number of differential privacy mechanisms.
Some mechanisms are designed to provide broad support for many types of queries~\cite{dwork2006calibrating, nissim2007smooth, mcsherry2007mechanism, mcsherry2009privacy, proserpio2014calibrating, mohan2012gupt, Blocki:2013:DPD:2422436.2422449, kotsogiannis2019privatesql}, while others are designed to produce maximal utility for a particular application~\cite{hay2009accurate,sala2011sharing,karwa2011private,kasiviswanathan2013analyzing,Chen:2013:RMT:2463676.2465304,hardt2012simple, li2014data, li2015matrix, zhang2014towards, xiao2012dpcube, cormode2012differentially, qardaji2013differentially, acs2012differentially, xu2013differentially,erlingsson2014rappor, DBLP:conf/sigmod/HayMMCZ16}.
While mechanisms adopt unique strategies for enforcing differential privacy in their target domain, they generally share a common set of design choices and algorithmic components. For example, many mechanisms require addition of Laplace noise to the result of the query.

\section{The \systemhead Architecture}
\label{sec:ipq}

This section presents the system architecture and advantages of \system, and compares it against existing architectures for differentially private analytics. We first describe the design goals motivating the \system architecture. Then, in Section~\ref{sec:existing_systems}, we describe the limitations of existing architectures preventing previous work from attaining these goals. Finally, Section~\ref{sec:chorus_arch} describes the novel architecture of \system and provides an overview of our approach.


\paragraph{Design Goals.} The design of \system is motivated by the desire to enforce differential privacy at the scale of real-world industrial deployments. To that end, \system has the following design goals:

\begin{itemize}[topsep=1mm,leftmargin=4mm]
\itemsep0.2mm
\item Process data using a DBMS, not a custom system
\item Support a broad range of privacy mechanisms
\item Integrate easily with existing data environments
\end{itemize}

\noindent As we will demonstrate in the next section, achieving these goals is challenging, and no existing system manages to achieve all three. We emphasize the importance of integration with an existing, highly-tuned database management system (DBMS)---such systems are the result of decades of research, and the massive scale of modern data warehouses is made possible only by leveraging these results. A custom-built system specific to differential privacy is unlikely to ever match the performance of a highly-tuned DBMS designed for big data.


\paragraph{Motivating example: bounded sum queries.}
Consider a simple example query over a table called \lstinline|trips| containing information about taxi trips. Suppose we want to return the sum of miles driven over all of the trips in the database. We might use a query like this:

\begin{lstlisting}
SELECT SUM(trip_distance) FROM trips
\end{lstlisting}

Satisfying differential privacy for this query is challenging, because there is no obvious bound on the global sensitivity of the \lstinline|SUM|. Two neighboring databases differ in only a single row, but that row may have any \emph{value}, and adding a row to the database increases the sum by the \emph{value} of an attribute in the new row. Without some upper bound on the attribute values rows can have, it is not possible to bound the sensitivity of the summation query.

The usual strategy for solving this problem is \emph{clipping}: we first \emph{enforce} a bound on the maximum distance of any trip in the database, then perform the sum on the clipped distances. We can implement this strategy using a revised query:

\begin{lstlisting}
SELECT SUM(max(0, min(100, trip_distance))) 
FROM trips
\end{lstlisting}

\noindent The revised query has a global sensitivity of 100, because all trip distances are clipped to lie between 0 and 100 miles. We can achieve differential privacy for the revised query by adding Laplace noise scaled to $\frac{100}{\epsilon}$~\cite{dwork2014algorithmic}.

\subsection{Existing Architectures}
\label{sec:existing_systems}

\begin{figure*}[!htb]
\centering
\footnotesize  \includegraphics[width=.95\linewidth]{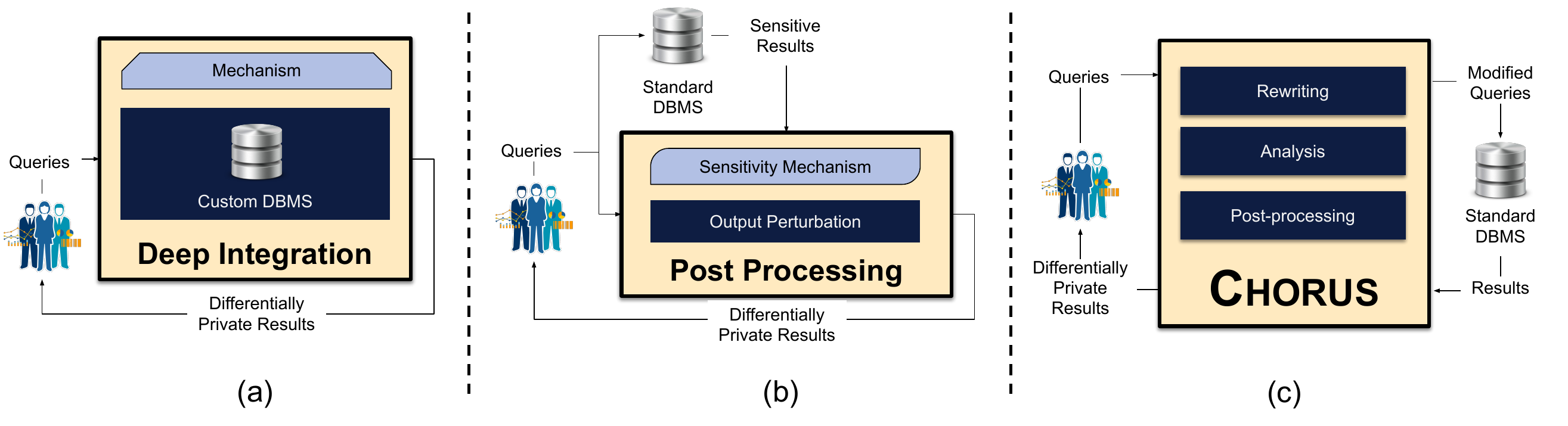}

  \begin{tabular}{l l l}
    \begin{minipage}{.29\textwidth}
      \raggedright
    \textbf{Pros:}
    \begin{itemize}
    \item Broad mechanism support
    \end{itemize}
    \textbf{Cons:}
    \begin{itemize}
    \item Poor scalability
    \item Higher software complexity
    \item New system required for each mechanism
    \end{itemize}
    \end{minipage}
    &
      \begin{minipage}{.32\textwidth}
        \raggedright
      \textbf{Pros:} 
      \begin{itemize}
      \item Good scalability
      \item DBMS-independent
      \end{itemize}
      \textbf{Cons:}
      \begin{itemize}
      \item Supports only post-processing mechanisms
      \end{itemize}
    \end{minipage}
    &
      \begin{minipage}{.3\textwidth}
        \raggedright
      \textbf{Pros:} 
      \begin{itemize}
      \item Good scalability
      \item DBMS-independent
      \item Broad mechanism support
      \end{itemize}
    \end{minipage}
      \\
  \end{tabular}

  \caption{Existing architectures: (a), (b); architecture of \system: (c).}
  \label{fig:sample_architectures}
  \rmfigspace
\end{figure*}

Existing systems for enforcing differential privacy for data analytics tasks adopt one of two architecture types: they are either \emph{deeply integrated} systems or \emph{post processing} systems. These architectures are summarized in Figure~\ref{fig:sample_architectures}(a) and Figure~\ref{fig:sample_architectures}(b).  PINQ~\cite{mcsherry2009privacy}, Weighted PINQ~\cite{proserpio2014calibrating}, GUPT~\cite{mohan2012gupt}, and Airavat~\cite{roy2010airavat} follow the \emph{deep integration} architecture: each one provides its own specialized DBMS, and cannot be used with a standard DBMS. As described earlier, the use of a specialized DBMS is likely to prevent the use of these systems for large-scale deployments.

The {\sc Flex}~\cite{allegro} system uses the \emph{post processing} architecture: it runs the analyst's original query on the database, then adds noise to the result.
This approach supports mechanisms that do not modify the semantics of the original query, like Elastic Sensitivity, PINQ, and Restricted sensitivity. The major advantage of the post processing architecture is that it is compatibile with existing DBMSs.

However, the post processing architecture is \emph{fundamentally incompatible} with mechanisms that change how the original query executes---such as queries that perform clipping, like the motivating example above. {\sc Flex} is not capable of answering \lstinline|SUM| queries with differential privacy unless bounds on the values of the summed columns are known \emph{a priori} (which is often not the case).

Many differential privacy mechanisms make even more complicated changes to the query the analyst actually wants to answer. For example, Sample \& Aggregate splits the database into chunks and runs the analyst's query on each chunk separately, then aggregates the results, and WPINQ assigns a weight to each row of the database and updates these weights as the query executes. More recent algorithms, like MWEM, the Matrix Mechanism, and others are even more complicated. These approaches are impossible to implement via post-processing: they require running queries which are \emph{different} from the analyst's original query, and they often require multi-stage \emph{interaction} with the DBMS.

The \emph{deeply integrated} and \emph{post processing} architectures in Figure~\ref{fig:sample_architectures}(a) and (b) therefore both fail to address two major challenges in implementing a practical system for differentially private data analytics:

\begin{itemize}[topsep=1mm,leftmargin=4mm]
\itemsep1.0mm
\item Deeply integrated systems use custom DBMSs, which are unlikely to achieve parity with mature DBMSs in terms of performance and scalability, query optimization, recoverability, and distribution.
\item Neither architecture supports all of the different mechanisms discussed earlier. The deeply integrated architecture requires building a new DBMS for each mechanism, while the post processing architecture is inherently incompatible with some mechanisms. 
\end{itemize}



\subsection{The \systemhead Architecture}
\label{sec:chorus_arch}

In \system, we propose a novel alternative, which we call the \emph{cooperative} architecture. As shown in Figure~\ref{fig:sample_architectures}(c), the cooperative architecture has two major differences with existing architectures:

\begin{itemize}[topsep=1mm,leftmargin=4mm]
\itemsep1.0mm
\item \system integrates tightly with an existing unmodified SQL DBMS, which holds the sensitive data.
\item \system can perform post-processing operations \emph{and} modify the way queries execute, enabling implementations of all the mechanisms discussed earlier.
\end{itemize}

\system provides a flexible framework for implementing differential privacy mechanisms in the cooperative architecture. In this architecture, the analyst specifies the task to be completed via one or more queries to be answered, plus additional metadata (for example, the desired values for the privacy parameter(s)). \system provides the following components:

\begin{itemize}[topsep=1mm,leftmargin=4mm]
\itemsep1.0mm
\item \textbf{Rewriting} supports generating new SQL queries, by modifying queries from the analyst's workload or generating new ones.
\item \textbf{Analysis} supports analyzing queries in the workload to determine their properties (e.g. sensitivity).
\item \textbf{Post-processing} supports post-processing of query results (e.g. to combine results or add noise).
\end{itemize}

\noindent For counting queries like those supported by {\sc Flex}~\cite{allegro}, \system can simply execute the queries in the workload and add noise to the results via post-processing. For a \lstinline|SUM| query requiring clipping, like our motivating example, \system rewrites the query to include the clipping bound specified by the analyst in the metadata, executes the rewritten query on the DBMS, and adds noise to the result. For more complicated mechanisms like the Matrix Mechanism~\cite{li2010optimizing}, \system may generate entirely new queries not present in the original workload.
\system has two key advantages over previous work:

\begin{itemize}[topsep=1mm,leftmargin=4mm]
\itemsep1.0mm
\item \system is DBMS-independent (unlike the deeply integrated approach): it requires neither modifying the database nor switching to purpose-built database engines. Our approach can therefore leverage existing high-performance DBMSs to scale to big data.
\item \system can implement a wide variety of
  privacy-preserving techniques. Unlike the post processing approach, \system is compatible with all of the mechanisms discussed earlier, and many more.

\end{itemize}


\system's architecture is specifically designed to be easily integrated into existing data environments. We report on the deployment of \system \extend{at Uber}{in a large technology company} in Section~\ref{sec:uber_deployment}.

\paragraph{Challenges.}
Developing differential privacy mechanisms to target the cooperative architecture is much more challenging than developing either deeply integrated or post-processing solutions. In particular, this model requires interacting closely with a DBMS which may have a non-standard dialect or feature set, analyzing SQL queries which may have complicated structure or target a specific dialect, and generating new queries which target the appropriate DBMS.

\system is designed specifically to address these challenges. We detail the solutions to each one in Section~\ref{sec:sum}. We also develop a number of case studies (in Section~\ref{sec:intr-priv-quer-1}) to demonstrate how the features of \system support programmers in developing differential privacy mechanisms in the cooperative architecture, following the \emph{rewrite-analyze-postprocess} structure outlined above.

\subsection{Threat Model}

Typical deployments of \system mechanisms involve three kinds of parties: \emph{data subjects}, who contribute sensitive data to the database, the \emph{data curator}, who manages the database containing the sensitive data, and \emph{analysts}, who submit queries to be answered using the sensitive data. \system is designed primarily to protect the sensitive data contributed by data subjects against malicious analysts.

As with most systems designed around the central model of differential privacy, we assume that the data curator behaves honestly. The DBMS, \system-based mechanisms, and other systems maintained by the data curator for answering queries are assumed to be trusted, and cannot be corrupted.

The adversary in this setting is represented by a group of one or more malicious analysts, who would like to discover a fact about an individual data subject in the sensitive data. The analyst may submit arbitrary queries to the system, designed to expose private information about an individual. These queries may be adaptively chosen based on previous results, and more than one analyst may collude to infer private information. \system is designed to guarantee that each query response satisfies differential privacy, which implies a bound on the total privacy cost of all queries posed by the analyst.


Our deployment enforces that the adversary may access the sensitive data \emph{only} via the centralized query interface. All components of the system, including the query interface, the privacy budget accountant, \system, and the DBMSs themselves, are protected from tampering by the adversary via access-control protections.


{\revisioncolor
\subsection{Selecting a Mechanism}

As described in Section~\ref{sec:backgr-diff-priv}, many  differential privacy mechanisms exist, and many of these require the analyst to re-phrase queries in a new way or provide additional inputs. As a result, it is not always possible to \emph{automatically} select the best mechanism for answering a SQL query posed by the analyst---the best mechanism might depend on domain knowledge, other queries in the workload, or the ability of the analyst to re-phrase the query.
For certain classes of queries---for example, linear queries over a single database table---Hay et al.~\cite{kotsogiannis2017pythia} demonstrate that a machine learning-based approach can leverage properties of the data to select a mechanism most likely to yield high utility.

The interface provided to the analyst in each deployment of \system will therefore depend on the analyst's expected expertise in differential privacy.

\begin{itemize}[topsep=1mm,leftmargin=4mm]
  \itemsep0.5mm
\item \emph{Non-experts} will submit standard SQL queries, without any knowledge about how differential privacy is being enforced. \system can attempt to select a mechanism which supports the features used in the query. If no suitable mechanism is found, \system rejects the query.
\item \emph{Privacy-conscious} analysts will submit workloads of queries written in subsets of SQL (e.g. linear queries), and \system can select the best approach (e.g. using a machine learning model).
\item \emph{Privacy experts} will manually select a mechanism, and phrase their queries appropriately for that mechanism (e.g. for the Sparse Vector Technique, a sequence of queries and a threshold).
\end{itemize}

\noindent The \system API is designed to facilitate all three possibilities. For our prototype deployment (discussed in Section~\ref{sec:uber_deployment}), we expected users to be non-experts, and implemented a simple rule-based mechanism to select a mechanism based on the aggregation function used and whether or not the query contained joins. Section~\ref{sec:budgeting-mechanism} describes the use of the \system API for this purpose.
}

\subsection{Privacy Budget Management}

{\revisioncolor
We have designed \system to be flexible in its handling of the privacy budget, since the best approach in a given setting is likely to depend on the domain and the kinds of queries posed.
}


\tocut{The simplest possible strategy for budget management is based on the standard composition theorem~\cite{dwork2006calibrating} for differential privacy. This is trivially implemented in \system by tracking the remaining budget, subtracting the budget requested for each processed query, and refusing to process new queries when the budget is exhausted.}

One approach to budget management involves tracking a single global budget, subtracting from the budget when each query runs using standard composition. One straightforward optimization is to use  \emph{advanced composition}~\cite{dwork2010boosting}, which improves the total budget for $k$ queries to be proportional to $\sqrt{k}$. Recent advances in composition, like R\'{e}nyi differential privacy~\cite{mironov2017renyi}, zero-concentrated differential privacy~\cite{bun2016concentrated}, and truncated concentrated differential privacy~\cite{bun2018composable}, can be directly applied in a similar way.

Some mechanisms build a differentially private synthetic representation of the data, and use this representation to answer queries (e.g. MWEM and the matrix mechanism). This is another form of budget management: once the representation is built, it can be used to answer an unbounded number of queries without incurring additional privacy cost. Such mechanisms often offer better accuracy over workloads of queries than any composition approach which supports online answering of queries.

{\revisioncolor
We describe the \system API for implementing budgeting strategies in Section~\ref{sec:budgeting-mechanism}. Mechanism definitions return the privacy cost of one execution, and approaches to budgeting can be built on top of this interface.
}

{\revisioncolor
\subsection{Assumptions \& Limitations}

\system's guarantees rely on the soundness of several underlying components. Bugs in these components could cause a failure of the guarantee, and the release of sensitive information without differential privacy.

\paragraph{Correctness of Underlying Libraries.} \system uses the Apache Calcite~\cite{calcite} framework for parsing SQL queries and translating them to a bag-based variant of relational algebra. A bug in Calcite could cause a query to be wrongly parsed or converted. Such a failure would result in incorrect results being returned to the analyst, but would not cause a failure in differential privacy, since \system analyzes, rewrites, and executes the \emph{final} output of Calcite's processing pipeline. In addition, Apache Calcite is widely used and therefore likely to be reliable.

\paragraph{Soundness of Abstract Interpretation.} To analyze a query's sensitivity, \system performs abstract interpretation of the query (see Section~\ref{sec:intr-priv-quer-1}). \system provides an abstract interpretation framework, which enables implementing many different kinds of analyses. A bug in either the framework or the implementation of a specific analysis could cause unsoundness in the analysis results, leading to a failure to ensure differential privacy. To mitigate this risk, we designed the framework to be compact and easily audited; further mitigation using formal verification techniques might also be possible in future work.

\paragraph{Semantics of DBMSs.} A more subtle failure of the abstract interpretation occurs when the concrete semantics of the DBMS used to execute queries do not match the semantics encoded by the abstract interpreter. This is a special concern for \system, since we aim to support many different DBMSs, and because DBMSs sometimes differ significantly in their semantics~\cite{guagliardo2017formal}. Mitigation of this risk would require formal analysis of specific DBMS implementations to verify their compliance with a standard semantics.

\paragraph{Dialects and Other Languages.} \system uses Apache Calcite~\cite{calcite} to parse and process queries, and works for queries that Calcite supports. Calcite does offer support for a number of SQL dialects, but it may not support all of the vendor-specific extensions offered by a particular DBMS, and \system will therefore not support them either (without modifications to Calcite). Code written in languages other than SQL (including stored or user-defined functions) are not supported for the same reason.

\paragraph{Sources of Randomness.} \system relies on randomness generated by both the DBMS and the Java runtime, and previous work has shown~\cite{mironov2012significance} that inadequate sources of randomness can lead to a failure of differential privacy. However, the same work also demonstrated simple solutions for improving faulty sources of randomness to recover differential privacy. Large-scale deployments of \system should verify that both the DBMS and Java runtime used provide high-quality sources of randomness, or implement the appropriate countermeasures.

\paragraph{Correctness of Mechanism Implementations.} Bugs may also exist in mechanism implementations themselves. The \system framework is designed to simplify mechanism implementations to reduce bugs, but it does not eliminate them entirely. Integrating an approach for formal verification of mechanism correctness (e.g.~\cite{reed2010distance, gaboardi2013linear, de2019probabilistic, zhang2019fuzzi, near2019duet, DBLP:journals/lmcs/BartheEHSS19, Barthe:POPL12,Barthe:TOPLAS:13, Barthe:LICS16, Sato:LICS19, DBLP:journals/pacmpl/AlbarghouthiH18, Barthe:CSF14, zhang2017lightdp, wang2019proving}) could ensure bug-free mechanisms, and is an exciting area for future work.
}

\section{The \system Programming Framework}
\label{sec:sum}

The \system framework provides a Scala library for implementing differential privacy mechanisms in the cooperative model we have proposed. The library provides support for all three components of the model:

\begin{itemize}[topsep=1mm,leftmargin=4mm]
\itemsep0.3mm
\item A \emph{rewriting} component, with support for modifying existing SQL queries and generating new ones
\item An \emph{analysis} component, which provides an abstract interpretation framework for analyzing SQL queries
\item A \emph{post-processing} component, which provides utilities for post-processing results
\end{itemize}

\noindent In this section, we describe each of the three components. We begin with our simple motivating example from earlier: a \lstinline|SUM| query over a column with no known upper bound. As described earlier, bounding the sensitivity of this query requires clipping the values being summed, but this is impossible to accomplish by post-processing alone.

We will implement a \system mechanism to answer such queries in three steps: (1) rewrite \lstinline|SUM| aggregations in the analyst's query to perform clipping; (2) analyze the rewritten query to determine its sensitivity; (3) run the rewritten query and add noise to the result based on the sensitivity computed in step (2).

\subsection{Rewriting}

For the first step, \system provides a powerful \emph{rewriting} API for modifying the analyst's queries. The following Scala code implements the rewriter for step (1) above.

\begin{lstlisting}[language=scala]
def rewriteClip(l: Double, u: Double, 
                root: Relation): Relation = {
  root.rewriteRecursive(UnitDomain) { 
    (node, orig, _) =>
      node match {
        case Relation(a: Aggregate) => {
          val r = a.mapCols { col => 
            max(l, min(u, col.expr)) AS col.alias }
          (r, ())
        }
        case _ => (node, ())
  } } }
\end{lstlisting}

\system's rewriter API contains an embedded DSL for building SQL queries; the expression \lstinline[language=scala]|max(l, min(u, col.expr))| uses this embedded DSL to generate a new expression for the argument to the \lstinline|SUM| aggregation. Running this rewriter on our simple query from earlier, with a lower bound of 0 and upper bound of 100, produces the following change:

\begin{lstlisting}
SELECT SUM(trip_distance) FROM trips
\end{lstlisting}
\mydownarrow
\begin{lstlisting}
SELECT SUM(max(0, min(100, trip_distance))) AS sum 
FROM trips
\end{lstlisting}

\system's rewriting library is designed to address the challenges of rewriting tasks like the one above. Solving such tasks requires matching on generalized patterns in the query, and replacing sections of the query with new text. This process requires building an abstract syntax tree for the query, since searching through its text (e.g. with regular expressions) will not provide enough information about the query's structure to implement the correct semantics. Modern DBMSs use a variety of SQL dialects; many SQL queries in production are hundreds or thousands of lines long and use many of these special features, so analyzing these queries is difficult.

\system provides a query parser based on Apache Calcite~\cite{calcite}, and also allows extending the parser to support features of specialized dialects. We translate special features into an abstract syntax tree (AST) based on relational algebra for rewriting, and provide an API for pattern matching on pieces of the AST. The \lstinline|rewriteRecursive| method, for example, makes it easy to search for a particular pattern in the query and replace it with something new. This method also allows the program to perform simultaneous analysis of the query, using the abstract interpretation framework described in the next section.

A second challenge of rewriting is producing \emph{new} queries (or sections of queries). \system provides an embedded domain-specific language (DSL) as a library of Scala operators for this purpose. SQL queries produced using our DSL have two major advantages over a simpler solution based on formatting strings: first, they are more readable, and can easily incorporate other AST nodes (as in \lstinline[language=scala]|min(u, col.expr)| above), and second, the AST objects produced by our DSL operators can be easily translated \emph{back} into different SQL dialects. For example, if the DBMS used for deployment provides the \lstinline|minimum| function instead of \lstinline|min|, \system can translate the above query appropriately for this dialect.

\subsection{Analysis}

The next step is to analyze the rewritten query to determine its sensitivity. For this purpose, \system provides a query analysis API which implements an abstract interpretation framework for SQL queries. {\revisioncolor Abstract interpretation is the execution of a program using \emph{abstract values} instead of concrete (normal) values. Abstract values are members of an \emph{abstract domain}, and each abstract value represents a set of possible concrete values. For example, the abstract domain $\{ \mathtt{Even}, \mathtt{Odd}, \mathtt{Unknown} \}$ might correspond to the concrete domain of the integers, and the abstract value $\mathtt{Even}$ represents the set of even numbers. Abstract interpretation enables analysis of program properties: if a program outputs the value $\mathtt{Even}$, then we know its output is always an even number. In \system, we use abstract interpretation on SQL queries to determine properties like the sensitivity of the query independent of the underlying concrete data.}

The sensitivity analysis for our rewritten query depends on three properties. First, the \emph{stability}~\cite{mcsherry2009privacy} of the underlying relation is 1. {\revisioncolor The stability of a relation measures how much the transformations used to create it from underlying tables change the number of rows before aggregation. A stability of 1 means that these transformations do not change the number of rows.} Second, each value is clipped to lie between 0 and 100 before being aggregated. Third, the aggregation being performed is a \lstinline|SUM|, whose global sensitivity is equal to $(u - l) \cdot s$ where $u$ and $l$ are the upper and lower clipping bounds, respectively, and $s$ is the stability of the underlying relation.

We can use \system to build an analysis which encodes these three properties by developing a pair of abstract domains that track properties of subexpressions of the query. {\revisioncolor The abstract values in these domains are placeholders for arbitrary relations, annotated with upper bounds on their stabilities or sensitivities.} The \emph{stability domain} tracks the stability (defined above) of relational expressions; the \emph{sensitivity domain} tracks upper and lower bounds of values, and the global sensitivity (defined in Section~\ref{sec:backgr-diff-priv}) of values.

\begin{lstlisting}[language=scala]
case class ColSens(
  sensitivity: Option[Double],
  upper: Double,
  lower: Double)

object SensDomain 
  extends AbstractDomain[ColSens] {
  override val bottom: ColSens =
    ColSens(
      sensitivity = None,
      upper = PositiveInfinity,
      lower = NegativeInfinity) }
\end{lstlisting}

The second part is to implement the analysis itself, which encodes the rules for updating the domain. For stability, we will return a stability of 1 for relations resulting from a single table, and a stability of infinity otherwise. For sensitivity, we will follow the rules outlined above: the sensitivity and upper and lower bounds of a column reference start out infinite; upper and lower bounds are introduced using the \lstinline|max| and \lstinline|min| functions, respectively; and sensitivity is bounded using the \lstinline|COUNT| and \lstinline|SUM| aggregation functions. The definition of the analysis appears in Figure~\ref{fig:analysis}.

\begin{figure}
\begin{lstlisting}[language=scala]
class SensAnalysis extends ColAnalysis(SensDomain) {
  override def transferAggregate(
      node: Aggregate,
      state: ColSens) = {
    node.aggFunction match {
      case Some(func) =>
        val newSensitivity = func match {
          case COUNT => state.nodeFact.stability
          case SUM   =>
            (state.nodeFact.upper - state.nodeFact.lower)
            * state.nodeFact.stability
          case _     => PositiveInfinity
        }

        colState.copy(
          sensitivity = Some(newSensitivity),
          upper = PositiveInfinity,
          lower = NegativeInfinity)
    } }

  override def transferExpression(
      node: RexNode,
      state: ColSens) = {
    val bot = state.copy(
      sensitivity = None,
      upper = PositiveInfinity,
      lower = PositiveInfinity)

    node match {
      case c: RexCall =>
        c.getOperator.getKind match {
          case SqlKind.Max =>
            state.copy(upper=c.getOperands[0])
          case SqlKind.Min =>
            state.copy(lower=c.getOperands[0])
          case _ => bot
        }
      case _ => bot
    } } }
\end{lstlisting}

  \caption{Static Analysis for Determining Sensitivity of \lstinline|SUM| and \lstinline|COUNT| Queries, with Clipping}
  \label{fig:analysis}
  \rmfigspace
  \rmfigspace
\end{figure}

The abstract interpretation framework in \system addresses a key challenge of deploying differentially private mechanisms for realistic SQL queries: SQL queries can be extremely complicated, and building sound analyses of these queries (e.g. for determining sensitivity) is correspondingly difficult.

The \system analysis library allows a small amount of code to define sound analyses which support these complicated queries. The analysis we have defined here can be used even on queries with features we have not discussed, including subqueries, queries with \lstinline|WHERE| clauses, and so on. This ability is based on the same parser and relational abstract syntax tree support described in the last section, extended with a dataflow analysis.

\system implements a dataflow library for relational ASTs based on the classic monotone framework~\cite{kam}. {\revisioncolor In this framework, the programmer defines an abstract domain as a lattice, provides a \emph{join} operator for computing the least upper bound of two points in the lattice, and writes \emph{transfer functions} to specify the effect of each operator in the programming language on its inputs. In \system, these are specified as regular Scala functions. The framework uses these to generate an abstract interpreter~\cite{nielson} that checks properties of programs by computing conservative approximations of the program's output.
}



This abstract interpretation framework greatly simplifies the task of analyzing a query to determine its properties. Above, we have defined a domain which tracks \emph{two} properties of the query: (1) bounded ranges for values in sub-parts of the query, introduced by clipping, and (2) sensitivity of the query, based on aggregation functions and the bounded ranges determined in part (1). Implemented directly on abstract syntax trees, such an analysis would comprise hundreds of lines of complicated code; \system's analysis library is designed to provide re-usable components to make these tasks simple.

\subsection{Post-Processing}

Finally, we use the rewriter and analyzer we have defined to produce a new SQL query with bounded sensitivity, run the query using the DBMS, and add Laplace noise to the results.

\begin{lstlisting}[language=scala]
def laplaceMechClip(query: String,
    epsilon: Double): List[Row] = {
  val rewritten = rewriteClip(0, 100)(query)
  val sens = SensAnalysis().analyze(rewritten)

  val r = DB.execute(rewritten)
  r + Utils.Laplace(sens / epsilon)
}
\end{lstlisting}

\noindent Here, the call to \lstinline[language=scala]{DB.execute} actually runs the query on the DBMS used in deployment. \system works with any DBMS with a JDBC interface, and the configuration information for the DBMS is specified in \system's configuration file.

This mechanism is ready for deployment using \system, which enables it to run alongside any standard SQL DBMS and scale to massive databases. Since the code we have defined here is part of a \emph{static} analysis of just the query, the scalability of our mechanism depends only on the ability of the cooperating DBMS to execute the rewritten query (which has only minor changes).

Our mechanism definitions are simply Scala functions, and can be exposed to analysts in a number of different ways depending on the deployment scenario (more in Section~\ref{sec:implementation}). The \system post-processing library provides a number of useful utilities, including the Laplace mechanism (used above), the Gaussian mechanism, the Exponential mechanism, and various forms of clipping.

This simple example illustrates how \system enables a \emph{rewrite-analyze-postprocess} pattern to achieve its design goals of enabling mechanisms which change how the query executes while integrating easily with an existing DBMS. We will see this pattern repeatedly in the more complicated mechanisms we will develop in Section~\ref{sec:intr-priv-quer-1}.

{\revisioncolor

  \subsection{Budgeting \& Mechanism Selection}
  \label{sec:budgeting-mechanism}

  The \system API provides two interfaces for privacy budget accounting: \lstinline|PrivacyCost|, to represent privacy costs, and \lstinline|PrivacyAccountant|, to track the total cost of composing many mechanisms.

  The \lstinline|PrivacyCost| interface requires the programmer to define the \lstinline|+| method in accordance with the sequential composition property of the corresponding privacy definition. The following two classes define privacy costs for pure $\epsilon$-differential privacy and R\'{e}nyi differential privacy~\cite{mironov2017renyi} (we have also defined privacy cost for $(\epsilon,\delta)$-differential privacy and Zero-concentrated differential privacy~\cite{bun2016concentrated}):

\begin{lstlisting}[language=scala]
case class EpsilonDPCost(epsilon: Double) 
    extends PrivacyCost {
  def +(other: PrivacyCost) = other match {
    case EpsilonDPCost(otherEpsilon) => 
      EpsilonDPCost(epsilon + otherEpsilon) }}

case class RenyiDPCost(alpha: Int, epsilon: Double) 
    extends PrivacyCost {
  def +(other: PrivacyCost) = other match {
    case RenyiDPCost(otherAlpha, otherEpsilon) =>
      RenyiDPCost(math.max(alpha, otherAlpha), 
                  epsilon + otherEpsilon) }}
\end{lstlisting}

  The \lstinline|PrivacyAccountant| class enables different approaches to computing the total budget used over many queries. The base class tracks the privacy costs of individual mechanisms, and the programmer defines a \lstinline|getTotalCost| method to compose these costs. For example, the following two classes define accountants  for advanced composition of pure $\epsilon$-differentially private mechanisms and for R\'{e}nyi differential privacy:

\begin{lstlisting}[language=scala]
class AdvancedCompositionAccountant(delta: Double) 
    extends PrivacyAccountant {
  def getTotalCost() = {
    val epsilons = costs.map(_.epsilon)
    val totalEpsilon = 2*(epsilons.max)*
      math.sqrt(2*(epsilons.length)*math.log(1/delta))
    EpsilonDeltaDPCost(totalEpsilon, delta) }}

class RenyiCompositionAccountant 
    extends PrivacyAccountant {
  def getTotalCost() = 
    costs.fold(RenyiDPCost(0, 0))(_ + _) }
\end{lstlisting}

  Finally, the \lstinline|ChorusMechanism| abstract class defines a standard interface for mechanisms, and integrates them with a chosen privacy accountant. Each mechanism class defines a \lstinline|run| method that returns a differentially private result and a \lstinline|PrivacyCost| object. To run a mechanism with accountant \lstinline|a|, the programmer calls \lstinline|execute(a)|, which invokes \lstinline|run|, adds the mechanism's cost to the accountant, and returns the result. For example, we can package our Laplace mechanism into a \system mechanism:

\begin{lstlisting}[language=scala]
class laplaceMechanism(epsilon: Double, l: Double, 
  u: Double, root: Relation) 
    extends ChorusMechanism[List[DB.Row]] {
  def run() = (laplaceMechClip(epsilon, l, u, root),
               EpsilonDPCost(epsilon)) }
\end{lstlisting}

These interfaces provide a flexible way to build systems that leverage existing mechanisms, permit building new mechanisms from old ones, and even allow implementing automatic mechanism selection (via ``mechanisms'' that use query properties to select from a list of mechanisms to run). For example, for our prototype deployment, we implemented a \system mechanism that runs one of three individual mechanisms using a simple rule-based approach.
}

\section{Mechanism Development with \system}
\label{sec:intr-priv-quer-1}



\newcommand{\init}{i}
\newcommand{\update}{j}
\newcommand{\countupdate}{c}
\newcommand{\arr}{\overset{\init,\update,\countupdate}{\rightarrow}}
\newcommand{\arrrep}{\overset{\Gamma, \Gamma'}{\rightarrow}}
\newcommand{\arrlap}{\overset{\gamma}{\rightarrow}}
\newcommand{\arrsamp}{\overset{\mathcal{A}}{\rightarrow}}
\newcommand{\meta}{\mbox{m}}
\newcommand{\agg}{\mathcal{A}}
\newcommand{\metu}{\mbox{u}}




This section demonstrates the use of \system to implement a number of different mechanisms, from simple ones based on the Laplace mechanism to more advanced algorithms like Sample and Aggregate and MWEM.

\subsection{Average Queries}

Average queries are typically answered with differential privacy by transforming the query into two separate differentially private queries---a \lstinline|SUM| and a \lstinline|COUNT|---and performing the division as post-processing.

We can implement this approach in \system by re-using the building blocks we have already built. First, we will develop two rewriters: one that turns \lstinline|AVG| into \lstinline|SUM|, and another that turns \lstinline|AVG| into \lstinline|COUNT|. We will rewrite the input query twice, once with each rewriter, and then run both rewritten queries using the mechanism we developed in Section~\ref{sec:sum}. The full implementation appears in Figure~\ref{fig:avg}.

This modular approach illustrates a key benefit of \system's design: the ability to re-use existing mechanisms to implement new ones. We will see this pattern used again in later mechanisms to reduce complexity.

\begin{figure}
\begin{lstlisting}[language=scala]
def avgMech(query: Relation, epsilon: Double) = {
  def replaceAvg(q: Relation, aggFn: Aggregate) = {
    Chorus.recursiveRewrite(q) { (node: Relation) =>
      node match {
        case Relation(a: SqlAvgAggFunction) => {
          val r = a.mapCols { col => 
            aggFn(col) AS col.alias }
          (r, ())
        }
        case _ => (node, ())
   } } }
  
  val sumQuery = replaceAvg(query, Sum)
  val countQuery = replaceAvg(query, Count)

  val r1 = laplaceMechClip(sumQuery, epsilon / 2)
  val r2 = laplaceMechClip(countQuery, epsilon / 2)
  r1 / r2 }
\end{lstlisting}
  \caption{\lstinline|AVG| Mechanism in \system}
  \label{fig:avg}
  \rmfigspace
\end{figure}

\subsection{Report Noisy Max}

The \emph{report noisy max} mechanism~\cite{dwork2014algorithmic} takes a list of queries as input, adds independently drawn noisy to each one, and returns the \emph{index} of the maximum noisy result. The key advantage of report noisy max is that it consumes privacy budget proportional to \emph{one query}, regardless of the length of the list of queries specified by the analyst.

This mechanism can be implemented in \system by extending the ideas in the \lstinline|laplaceMechClip| mechanism to a list of queries, and returning only the index of the maximum value in the resulting list:

\begin{lstlisting}[language=scala]
def reportNoisyMax(queries: List[Relation], 
  epsilon: Double): Int = {
    val results = queries.map(
      laplaceMechClip(_, epsilon, 0, 1))
    Utils.argmax(results) }
\end{lstlisting}


Our implementation of the report noisy max mechanism demonstrates two important principles. First is the use of a workload of queries: since \system mechanisms are implemented as regular Scala functions, they can accept queries in any format, including a workload of SQL queries. Second is the splitting of mechanism logic between Scala and SQL: query results can be post-processed with Scala code in arbitrary ways (here, we use Scala to find the maximum workload result). 

\subsection{Exponential Mechanism}

The \emph{exponential mechanism}~\cite{mcsherry2007mechanism} is the generalized version of report noisy max: it selects an element of a set $R$ which approximately maximizes the value of a \emph{scoring function}. The scoring function $q : D^n \rightarrow R \rightarrow \mathbb{R}$ assigns numeric scores to each element of $R$ based on the private database (in report noisy max, the scoring function simply returns the value of the query). For a scoring function with sensitivity $\Delta = GS(q)$ and a database $X$, the exponential mechanism outputs $r \sim R$ with probability proportional to:
\[\text{Pr}[r] \sim \exp\Big(\frac{\epsilon q(X, r)}{2\Delta}\Big)\]
In report noisy max, the scoring function was fixed, so it was easy to assume something about its sensitivity and apply it directly to each query in the workload. The generalized exponential mechanism makes this process more difficult: since the analyst specifies the scoring function, we cannot assume anything about its sensitivity.

The solution is to ask the analyst to provide the scoring function \emph{as a query}, so that we can analyze its sensitivity directly. This query should produce a relation mapping elements in $R$ to their scores. For example, the following query implements a scoring function which will allow selecting the day of the week with approximately the largest number of trips:

\begin{lstlisting}
SELECT day, COUNT(*) as score
FROM trips
GROUP BY day
\end{lstlisting}

\noindent We can use the same basic analysis as before to show that this query has a sensitivity of 1, and then perform the selection step of the exponential mechanism in a post-processing step.

\begin{lstlisting}[language=scala]
def exponentialMech(scoring: Relation, 
                    epsilon: Double) = {
  val s = SensAnalysis.analyze(scoring)
  val scores = DB.execute(scoring)
  val totalScore = scores.map(_._2).sum()
  val probabilities = scores.map { r =>
    (r(0), 
     (epsilon * (r(1) / totalScore)) / (2*s) ) }

  Utils.chooseWithProbability(probabilities)._1
}
\end{lstlisting}

\noindent Our implementation here uses \lstinline|Utils.chooseWithProbability| to select from the elements of $R$ with the appropriate probabilities (this function expects a list of tuples mapping domain elements to their probabilities, and implements weighted random selection). We return just the identity of the element which was selected, similar to the report noisy max mechanism.

\subsection{Sparse Vector Technique}

The \emph{sparse vector technique} (SVT)~\cite{dwork2014algorithmic} releases the index (but not the result) of the first query in a sequence of queries whose result exceeds a threshold set by the analyst. Like report noisy max, SVT consumes privacy budget proportional to just one query. In situations where only a small number of queries are likely to have large enough results to be useful to the analyst (but the analyst does not know which ones), SVT can be applied repeatedly to answer the useful queries while minimizing privacy budget consumption.

We can define a \system mechanism to implement SVT in a similar way to the report noisy max mechanism. In contrast to the report noisy max mechanism, however, SVT is \emph{iterative}---it runs the queries in the workload in sequence, and may halt before running all of them. SVT first adds Laplace noise to the threshold, then compares the noisy result of each query to the noisy threshold. SVT releases the index of the first query whose noisy value exceeds the noisy threshold.

\begin{figure}
\begin{lstlisting}[language=scala]
def sparseVectorMech(queries: List[Relation],
    threshold: Double, epsilon: Double) = {
  val sens = rewritten.map(SensAnalysis().analyze(_))

  // require sensitivity-1 queries
  if (sens.exists(_ > 1))
    return None

  // generate noisy threshold
  val T = threshold + Utils.laplaceSample(2/epsilon)

  for (i <- 0 to queries.length()) {
    val r = DB.execute(queries(i))
    if (r + Utils.Laplace(4/epsilon) >= T)
      return Some(i)  }

  return None }
\end{lstlisting}

  \caption{Sparse Vector Technique in \system}
  \label{fig:svt}
    \rmfigspace
\end{figure}

Our implementation appears in Figure~\ref{fig:svt}. First, it analyzes each query to ensure that the query's sensitivity does not exceed 1, and returns \lstinline[language=scala]|None| if not. Then, it uses \lstinline[language=scala]|Utils.Laplace| to add noise to both the threshold and each query result, and returns \lstinline[language=scala]|Some(i)| for the first index \lstinline[language=scala]|i| whose noisy query result exceeds the noisy threshold. If no query exceeds the threshold, we return \lstinline[language=scala]|None|.

This implementation corresponds to the \lstinline|AboveThreshold| algorithm described by Dwork and Roth~\cite{dwork2014algorithmic}. This algorithm can be combined with our earlier mechanisms to release the \emph{value} of the first query above the threshold:

\begin{lstlisting}[language=scala]
def sparseVectorMechValue(queries: List[Relation],
    thresh: Double, epsilon: Double) = {
  val i = sparseVectorMech(queries, thresh, epsilon/2)
  laplaceMechClip(queries(i), epsilon/2) }
\end{lstlisting}

\noindent This mechanism combines SVT with the Laplace mechanism to find the value of the first query above the threshold (not just its index), splitting the privacy budget between these two tasks. This mechanism illustrates the ease of combining mechanisms in \system to build complex functionality.

\subsection{Sample \& Aggregate}

The Sample \& Aggregate~\cite{nissim2007smooth,smith2011privacy}
mechanism works for all statistical estimators, but does not support
joins. Sample \& Aggregate has been implemented in
GUPT~\cite{mohan2012gupt}, a standalone data processing engine that
operates on Python programs, but has never been integrated with a
high-performance DBMS. As defined by Smith~\cite{smith2011privacy},
the mechanism has three steps:

\begin{enumerate}[topsep=1mm]
\item Split the database into disjoint \emph{subsamples}
\item Run the query on each subsample independently
\item Aggregate the results using a differentially private algorithm
\end{enumerate}


\noindent We can use the DBMS to accomplish tasks 1 and 2 by modifying
the analyst's original query. We add a \lstinline|GROUP BY| clause to
the original query which groups rows according to their row
number. Sample and aggregate does not require the subsamples to be
randomized, so basing the selection of the subsamples on row number
satisfies its requirements. For example, we can transform a simple
\lstinline|AVG| query as follows:

\begin{lstlisting}
SELECT AVG(trip_distance) FROM trips
\end{lstlisting}
\mydownarrow
\begin{lstlisting}
SELECT AVG(trip_distance), ROW_NUM() MOD (*@ n @*) AS _grp
FROM trips
GROUP BY _grp
\end{lstlisting}

\noindent This transformation generates $n$ subsamples and runs the
original query on each one. Once we have obtained the answers to query
on the subsamples, we can perform differentially private aggregation
as a post-processing step---with clipping followed by a noisy
average. The complete implementation appears in Figure~\ref{fig:saa}.

\begin{figure}
\begin{lstlisting}[language=scala]
def rewriteSAA(n: Int, root: Relation) = {
  root.rewriteRecursive(UnitDomain) { 
    (node, orig, _) =>
      node match {
        // Add new subsample number column
        case Relation(t: TableScan) => 
          (node.project(*, RowNumMod AS "_grp"), ())

        // Group by the subsample number
        case Relation(a: Aggregate) => 
          (a addGroupedColumn col("_grp"), ())

        case _ => (node, ())
  } } }

def saaMech(query: Relation, l: Double, u: Double, 
    numSubsamples: Int, epsilon: Double) = {
  // rewrite the query to perform subsampling
  val rewritten = rewriteSAA(numSubsamples, query)
  // execute rewritten query and get subsample results
  val r = db.execute(rewritten).map(_._2)

  // calculate sensitivity
  val sens = (u - l) / numSubsamples

  // calculate noisy average via clipping
  val mean = Utils.mean(Utils.clip(r, l, u))
  r + Utils.Laplace(sens / epsilon) }
\end{lstlisting}

  \caption{Sample \& Aggregate in \system}
  \label{fig:saa}
   \rmfigspace
\end{figure}

\subsection{Additional Mechanisms}

We present \system implementations of three additional mechanisms in in Appendix~\ref{sec:addit-mech}: Weighted PINQ, the Matrix Mechanism, and MWEM.

\section{Implementation \& Deployment}
\label{sec:implementation}
\label{sec:uber_deployment}

This section describes the deployment of \system to protect sensitive data and provide a secure, differentially private interface for analysts to query that data. We also describe our experience deploying \system to enforce differential privacy at \extend{Uber}{a large technology company}.


A typical deployment of a \system mechanism includes a centralized
query interface which allows the DBMSs containing sensitive data to be
queried \emph{only} via the mechanism. In such a deployment, untrusted
analysts submit queries to the centralized query interface, which runs
the mechanism and updates the privacy budget. The interface may enable
auditing of the budget by a trusted curator of the system, and
rejection of queries after the budget has been exhausted.

\paragraph{Implementation.}
Our implementation is built on Apache Calcite~\cite{calcite}, a generic query optimization framework that transforms input queries into a relational algebra tree and provides facilities for transforming the tree and emitting a new SQL query.
We built \system's custom dataflow analysis and rewriting components on Calcite to support the \system programming framework. The framework, mechanism-specific analyses, and rewriting rules are implemented in 5,096 lines of Java and Scala code. 

The approach could also be implemented with other query optimization frameworks or rule-based query rewriters such as Starburst~\cite{starburst}, ORCA~\cite{GPORCA}, and Cascades~\cite{cascades}.







\paragraph{Real-world Deployment.}
\system has been tested in a deployment to enforce differential privacy for queries over customer data at \extend{Uber}{a large technology company}. The primary goals of this deployment are to protect the privacy of customers from insider attacks, and to ensure compliance with the requirements of Europe's General Data Protection Regulation (GDPR)~\cite{gdpr}. The mode of deployment calls for \system to process more than 10,000 queries per day. 


The deployment's data environment consists of several DBMSs (three primary databases, plus several more for specific applications), and a single central query interface through which all queries are submitted. The query interface is implemented as a microservice that performs query processing and then submits the query to the appropriate DBMS and returns the results.

Deployment involved building a minimal wrapper around the \system library to expose its rewriting functionality as a microservice. The only required change to the data environment was a single modification to the query interface, to submit queries to the \system microservice for rewriting before execution. The wrapper around \system also queries a policy microservice to determine the security and privacy policy for the user submitting the query. This policy informs which mechanism is used---by default, differential privacy is required, but for some privileged users performing specific business tasks, differential privacy is only used for older data.

A major challenge of this deployment has been supporting the variety of SQL dialects used by the various DBMSs. This challenge motivated the built-in support for different dialects in the \system framework.

The privacy budget is managed by the microservice wrapper around \system. The microservice maintains a small amount of state to keep track of the current cumulative privacy cost of all queries submitted so far, and updates this state when a new query is submitted.
The current design of the \system microservice maintains a single global budget.

\section{Evaluation}
\label{sec:evaluation}

In this section, we evaluate the performance overhead of enforcing differential privacy using \system in the context of real-world queries on a large production dataset. {\revisioncolor Our results demonstrate that the mechanism implementations we have developed using \system scale effectively to realistic datasets using commodity DBMSs.}

{\revisioncolor
  Our evaluation is intended to demonstrate the scalability of the \system approach, rather than the ability of the mechanisms we implemented to produce accurate results. The accuracy we obtain in each of our experiments is a direct result of the underlying mechanism used in the experiment, and the same accuracy would be obtained using an alternative implementation of the same mechanism.}






\paragraph{Corpus.}
We use a corpus of \numqueries real-world queries containing all statistical queries executed by data analysts at \extend{Uber}{a large technology company} during a single month\tocut{October 2016}.\tocut{\footnote{We omit queries that return individual trips or users since differential privacy does not aim to provide high utility for such queries.}}
The corpus includes queries written for several use cases including fraud detection, marketing, business intelligence and general data exploration. 

\paragraph{Dataset.}
We used a database of data sampled from the production database in our evaluation. This database contained 300 million records representing trip data similar in nature to the New York City Taxi Trip Data~\cite{taxidata}---including information about trips, riders, and drivers.

\paragraph{Mechanisms.}
In our evaluation, we considered two more complicated variants of the sensitivity-based mechanism we developed in Section~\ref{sec:sum}: elastic sensitivity~\cite{allegro} and restricted sensitivity~\cite{Blocki:2013:DPD:2422436.2422449}. These mechanisms can answer existing counting and summation queries written by analysts unfamiliar with differential privacy, like those in our corpus. We also considered the performance of Weighted PINQ~\cite{proserpio2014calibrating}, since it performs more serious modifications to the analyst's query.

\subsection{Performance Overhead}
\label{sec:perf-overh-intr}

\begin{figure*}
  \centering
  \begin{tabular}{c c c}
    \includegraphics[width=.31\textwidth]{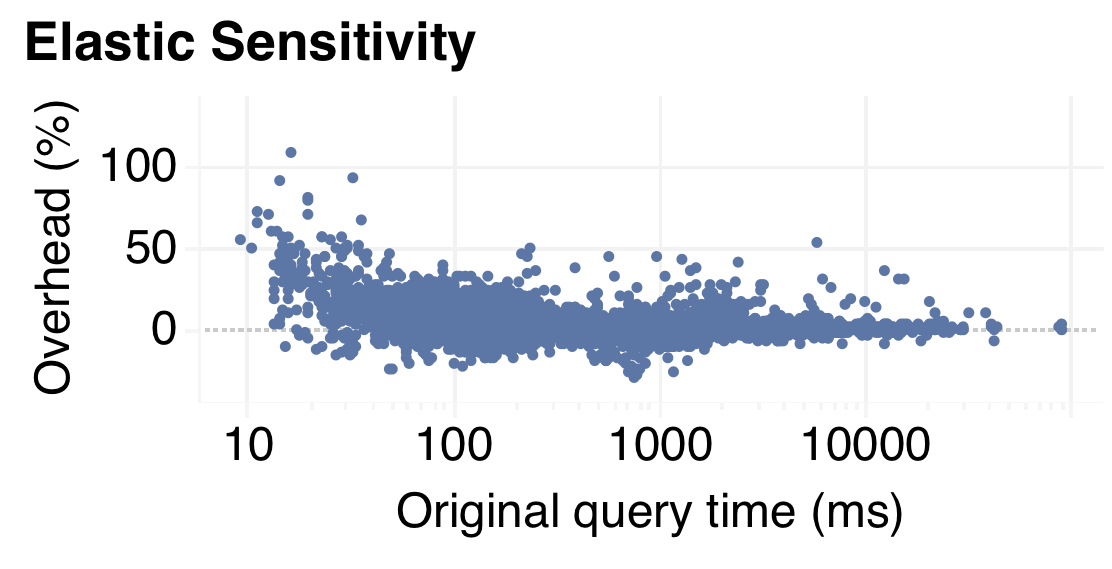} &
    \includegraphics[width=.31\textwidth]{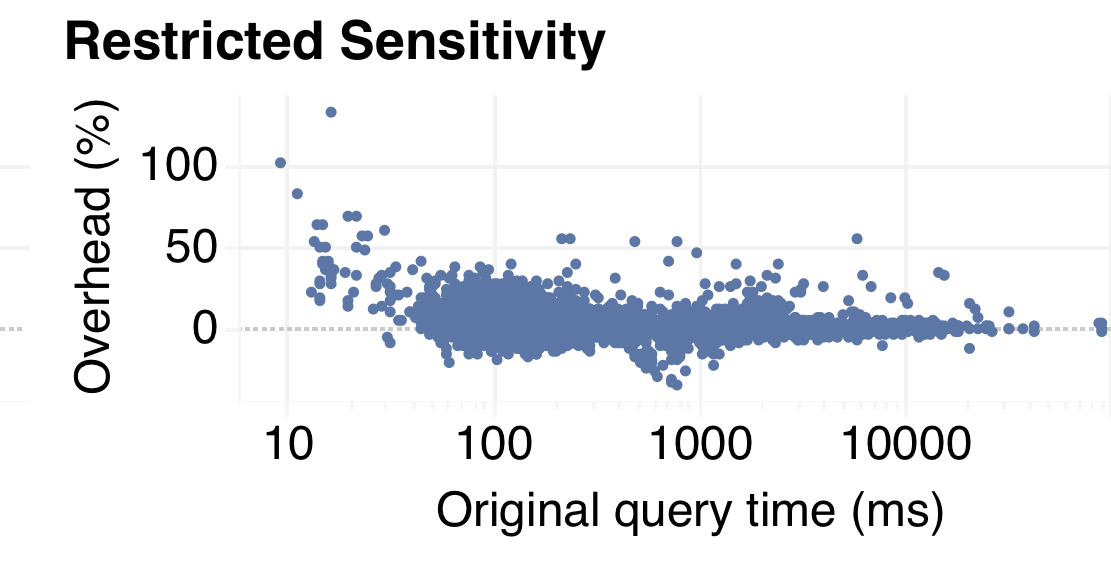} &
    \includegraphics[width=.31\textwidth]{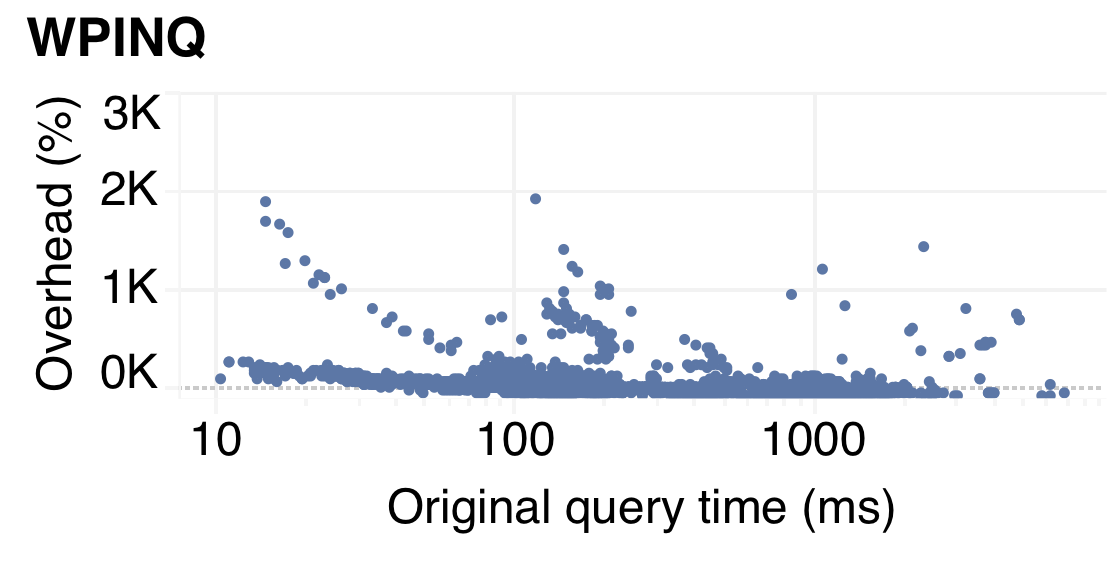} \\
  \end{tabular}
                                                                  \caption{Performance overhead of differential privacy mechanisms by execution time of original query.}
  \label{fig:performance_plots}
\end{figure*}




We conduct a performance evaluation demonstrating the performance overhead of several mechanisms implemented with \system.

\paragraph{Experiment Setup.}
We used a single HP Vertica 7.2.3~\cite{vertica} node containing 300 million records including trips, rider and driver information and other associated data stored across 8 tables. We submitted the queries locally and ran queries sequentially to avoid any effects from network latency and concurrent workloads.

To establish a baseline we ran each original query 10 times and recorded the average after dropping the lowest and highest times to control for outliers. Then, we ran each \system mechanism 10 times and recorded the average execution time, again dropping the fastest and slowest times. We calculate the overhead for each query by comparing the average runtime of the original query and the \system mechanism.

\paragraph{Results.}
Figure~\ref{fig:performance_plots} shows the distribution of overhead as a function of original query execution time. This distribution shows that the percentage overhead is highest when the original query was very fast (less than 100ms). This is because even a small incremental performance cost is fractionally larger for these queries.


WPINQ significantly alters the way the query executes (see Section~\ref{sec:intr-priv-quer-1}) and these changes increase query execution time. In particular, the query transformation adds a new join to the query each time weights are rescaled (i.e. one new join for each join in the original query), and these new joins result in the additional overhead. 
Figure~\ref{fig:performance_plots} shows that, in both cases, the performance impact is amortized over higher query execution times, resulting in a lower relative overhead for more expensive queries.


\subsection{Utility}
\label{sec:util-select-mech}

\begin{figure*}
  \centering
  \includegraphics[width=0.95\textwidth]{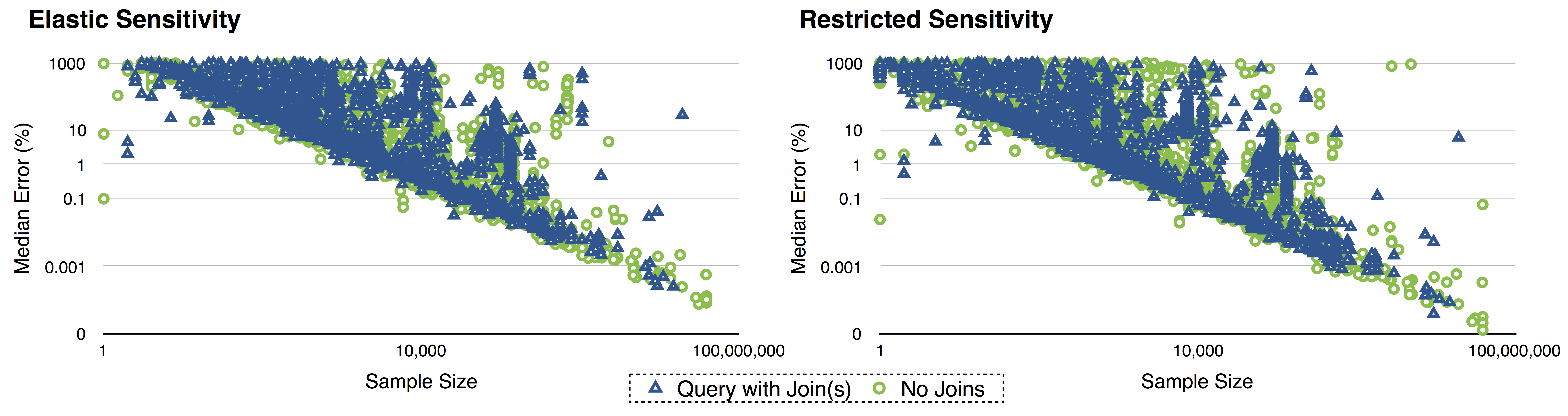}

  \caption{Utility of elastic sensitivity and restricted sensitivity, by presence of joins.}
  \label{fig:scatter_plots}
\end{figure*}

We also measured the ability of some of the mechanisms we have implemented with \system to produce accurate differentially private results for queries in our corpus, primarily as a study of whether or not differential privacy is a good fit for these queries. {\revisioncolor These mechanisms were previously proposed, and their accuracy properties were previously known. Our experimental results confirm existing knowledge using real-world queries and data.}

For this experiment, we used the elastic sensitivity and restricted sensitivity mechanisms described earlier. Evaluating the utility of more complicated mechanisms would require re-formulating the queries in our corpus with the help of the original analyst. All of these mechanisms have been previously evaluated on synthetic query workloads, and their ability to improve utility is well-understood; as we gain experience with practical deployments, analysts will begin to adopt these mechanisms and re-formulate their queries.

\paragraph{Experiment Setup.}
We use the same setup described in the previous section. For each query, we set the privacy budget $\epsilon=0.1$ for all mechanisms. For Elastic Sensitivity, we set $\delta = \frac{1}{n^2}$ (where $n$ is the database size).

We ran each query 10 times on the database and report the median relative error across these executions. For each run we report the relative error as the percentage difference between the differentially private result and the original non-private result. Consistent with previous evaluations of differential privacy~\cite{DBLP:conf/sigmod/HayMMCZ16} we report error as a proxy for utility since data analysts are primarily concerned with accuracy of results.

      
\paragraph{Query Sample Size.}
Our corpus includes queries covering a broad spectrum of use cases, from highly selective analytics (e.g., trips in San Francisco completed in the past hour) to statistics of large populations (e.g., all trips in the US).
Differential privacy generally requires the addition of more noise to highly selective queries than to queries over large populations, since the influence of any individual's data diminishes as population size increases. Consequently, a query's selectivity is important for interpreting the relative error introduced by differential privacy.
To measure the selectivity we calculate the \emph{sample size} of every aggregation function in the original query, which represents the number of input records to which the function was applied. 


\paragraph{Results.}
Figure~\ref{fig:scatter_plots} shows the results of this experiment. Both mechanisms exhibit the expected inverse relationship between sample size and error; moreover, this trend is apparent for queries with and without joins.

Importantly, a very large portion of the queries with large sample sizes have very small relative error---the majority of queries with a sample size of over 10,000, for example, have less than 1\% error. This suggests that for at least a subset of the queries in this production corpus, differential privacy can be enforced without significant harm to accuracy. Furthermore, more advanced mechanisms like the ones described in Section~\ref{sec:intr-priv-quer-1} could provide significant improvements for the accuracy of these queries.

A significant number of queries with small sample size also result in large relative error. It is likely that many of these queries are \emph{intended} to violate privacy---perhaps they examine the data of an individual or small set of individuals directly---and so differential privacy is probably not appropriate for these queries.

\subsection{Discussion and Key Takeaways}
\label{sec:disc-key-take}

\paragraph{Strengths \& weaknesses of differential privacy.}
The mechanisms we studied generally worked best for statistical queries over large populations. None of the mechanisms was able to provide accurate results (e.g. within 1\% error) for a significant number of queries over populations smaller than 1,000. These results confirm the existing wisdom that differential privacy is ill-suited for queries with small sample sizes. 
For large populations (e.g. more than 10,000), accurate differentially private results for real-world queries appears to be an achievable goal. A large set of such queries exists in our corpus. These results suggest that differential privacy can provide both strong privacy guarantees and accurate query responses for a large portion of the queries written by analysts in practice.

{\revisioncolor Our results also agree with the prior knowledge that queries with joins make ensuring differential privacy more challenging. For both mechanisms we used in our evaluation, the proportion of queries with less than 1\% error was much smaller for queries with joins than for queries without joins. Differential privacy for joins is an active area of research~\cite{allegro, kotsogiannis2019privatesql}, and we hope that future mechanisms can be implemented in \system to provide more accurate answers for these queries.}

\paragraph{Mechanism performance.}
%
%
%
Our performance evaluation demonstrates the scalability of mechanisms implemented with \system---the vast majority of the queries executed with the elastic sensitivity and restricted sensitivity mechanisms resulted in less than 50\% overhead, and the mean overhead for both was below 25\%. However, the results also highlight the variability in computation costs of differential privacy mechanisms---WPINQ's added joins resulted in high performance overhead for some queries.

\tocut{%
\paragraph{Performance impact of database engine.}
The performance of intrinsically private queries can depend on the database engine and transformations applied to the query. In this work we evaluate performance using Vertica, the primary database of our target environment. Our results provide a relative comparison of the mechanisms which we believe will hold generally, though the precise runtimes may differ on other database systems.

In this work we do not attempt to optimize the rewritten queries for performance\tocut{, although most query transformation frameworks provide a built-in suite of optimizations that could be invoked during transformation}. Since database engines employ highly customized query execution strategies, any such optimizations may not generalize to other databases. We leave as future work an investigation into optimizations and performance of intrinsically private queries on other databases.}





\section{Related Work}
\label{sec:related-work}

\paragraph{Differential Privacy.}
Differential privacy was originally proposed by Dwork~\cite{dworkdifferential2006, dwork2006calibrating, dwork2008differential}. The reference by Dwork and Roth~\cite{dwork2014algorithmic} provides an overview of the field.
%
%
%

Much recent work has focused on task-specific mechanisms for graph analysis~\cite{hay2009accurate,sala2011sharing,karwa2011private,kasiviswanathan2013analyzing,Chen:2013:RMT:2463676.2465304}, range queries~\cite{hardt2012simple, li2014data, li2010optimizing, li2015matrix, zhang2014towards, xiao2012dpcube, cormode2012differentially, qardaji2013differentially, acs2012differentially, xu2013differentially, mckenna2018optimizing}, and analysis of data streams~\cite{dwork2010differential,shi2011privacy}. As described in Section~\ref{sec:disc-key-take}, such mechanisms are complementary to our approach, and could be implemented on top of \system to provide more efficient use of the privacy budget.


\paragraph{Differential Privacy Systems.}
As differential privacy is more widely adopted, scalable implementations of differential privacy mechanisms have received more attention. Wilson et al. recently developed an open source, highly performant C++ library~\cite{wilson2019differentially} which provides basic differential privacy mechanisms as an extension to PostgreSQL. However, their approach does not support other DBMSs. It also does not provide a framework for development of additional mechanisms; more complex mechanisms, like MWEM, would need to be added directly to the C++ library as new primitives.

The problem of answering SQL queries has also received considerable attention in recent years. The {\sc Flex}~\cite{allegro} system answers counting queries with differential privacy, and since it implements the post-processing architecture, it is DBMS-independent. However, {\sc Flex} cannot implement more complex mechanisms like MWEM. PrivateSQL~\cite{kotsogiannis2019privatesql} answers SQL queries with differential privacy by generating differentially private \emph{synopses} of views, then using the synopses to answer queries. This approach is potentially scalable, but limits the set of implementable mechanisms.

A number of other systems for enforcing differential privacy have been developed. PINQ~\cite{mcsherry2009privacy} supports a LINQ-based query language, and implements the Laplace mechanism with a measure of global sensitivity. Weighted PINQ~\cite{proserpio2014calibrating} extends PINQ to weighted datasets, and implements a specialized mechanism for that setting.

Airavat~\cite{roy2010airavat} enforces differential privacy for MapReduce programs using the Laplace mechanism. Fuzz~\cite{gaboardi2013linear,haeberlen2011differential} enforces differential privacy for functional programs, using the Laplace mechanism in an approach similar to PINQ.
DJoin~\cite{narayan2012djoin} enforces differential privacy for queries over distributed datasets. Due to the additional restrictions associated with this setting, DJoin requires the use of special cryptographic functions during query execution so is incompatible with existing databases. GUPT~\cite{mohan2012gupt} implements the Sample \& Aggregate framework for Python programs. None of these systems offers integration with high-performance DBMSs.



\paragraph{Security \& Privacy via Query Rewriting.}
Automated query transformations have been used in previous work to implement access control.
Stonebreaker and Wong~\cite{stonebraker1974access} presented the first approach. Barker and Rosenthal~\cite{barker2002flexible} extended the approach to role-based access control by first constructing a view that encodes the access control policy, then rewriting input queries to add \lstinline|WHERE| clauses that query the view. Byun and Li~\cite{byun2008purpose} use a similar approach to enforce purpose-based access control: purposes are attached to data in the database, then queries are modified to enforce purpose restrictions drawn from a policy.

{\revisioncolor
Since then, a number of rewriting-based approaches have been proposed for enforcing access control. Agrawal et al.~\cite{agrawal2005extending} use query rewriting to enforce row-level privacy policies focused on access control. Bender et al.~\cite{bender2014explainable, bender2013fine} combine query rewriting with specially-designed views to enforce privacy policies organized in ``disclosure lattices.'' Wang et al.~\cite{wang2007correctness} propose fine-grained cell-level access control policies, with an enforcement mechanism based on query rewriting. Rizvi et al.~\cite{rizvi2004extending} propose an access control mechanism based on ``authorization views'' which define policies, and use query rewriting to re-phrase queries in terms of these views. Oracle's Virtual Private Database~\cite{browder2002virtual} enforces fine-grained access control policies using query rewriting. All of these are primarily focused on access control policies.

Mehta et al. present Qapla~\cite{mehta2017qapla}, a system which uses query rewriting to enforce policies written in SQL. Notably, Qapla includes \emph{aggregation policies}, which go beyond traditional access control policies to allow the release of aggregate statistics while protecting the underlying rows, but Qapla does not support differential privacy or other formal notions of privacy.

Guarnieri et al.~\cite{guarnieri2014optimal, guarnieri2016strong} explore the challenges of enforcing access control policies in the context of a complicated and expressive query language like SQL, and highlight the need for provable guarantees about the enforcement mechanism. Zhang and Mendelzon~\cite{zhang2005authorization} study one of these challanges---the problem of ``query containment''---in the context of proving correctness for query rewriting enforcement mechanisms. These results reinforce the value of additional work in the future to verify the correctness of our rewriting algorithms.
}

\section{Conclusion}

This paper presents \system, a framework which enables a novel cooperative architecture for enforcing differential privacy. \system works closely with a high-performance DBMS to scale differential privacy mechanisms to real-world deployments. \system combines the strengths of integrated implementations (whose scalability does not match high-performance industrial DBMSs) and post-processing based implementations (which scale up, but are incompatible with many modern differential privacy mechanisms).
%
We have described how \system can be deployed to provide differential privacy, and released it as open source~\extend{\cite{chorusdownload}}{\cite{chorusdownloadanon}}.





%

\ifextended
\section*{Acknowledgments}

\noindent The authors would like to thank Om Thakkar and the anonymous reviewers for their
helpful comments, and Uber's Privacy Engineering team for collaboration on this project.
This work was supported by the Center for Long-Term Cybersecurity, and DARPA \& SPAWAR under contract N66001-15-C-4066. The U.S. Government is authorized to reproduce and distribute reprints for Governmental purposes not withstanding any copyright notation thereon. The views, opinions, and/or findings expressed are those of the author(s) and should not be interpreted as representing the official views or policies of the Department of Defense or the U.S. Government.
\fi

\bibliographystyle{plain}
\bibliography{refs,allegro-refs,lang-refs}




\appendices
\section{Additional Mechanisms}
\label{sec:addit-mech}

This appendix presents \system implementations of three additional mechanisms: Weighted PINQ (Appendix~\ref{sec:weighted-pinq}), the Matrix Mechanism (Appendix~\ref{sec:matrix-mechanism}), and MWEM (Appendix~\ref{sec:mult-weights-mwem}).

\subsection{Weighted PINQ}
\label{sec:weighted-pinq}

Weighted PINQ (WPINQ) enforces differential privacy for 
counting queries with equijoins. A key distinction of this mechanism
is that it produces a differentially private \emph{metric} (called a \emph{weight}),
rather than a count. These weights are suitable for use in a workflow that generates differentially private
synthetic data, from which counts are easily derived.
The workflow described in~\cite{proserpio2014calibrating}
uses weights as input to a Markov chain Monte Carlo (MCMC) process.

Our \system implementation of WPINQ computes noisy weights for a given
counting query according to the mechanism's definition~\cite{proserpio2014calibrating}.
Since the weights are differentially private, they can be released to the analyst for use
with any desired workflow.


The WPINQ mechanism adds a weight to each row of the database,
updates the weights as the query executes to ensure that the query has
a sensitivity of 1, and uses the Laplace mechanism to add noise to the
weighted query result.
WPINQ has been implemented as a standalone data processing engine with
a specialized query language, but since the mechanism cannot be implemented via post-processing alone, it has not been integrated into any SQL DBMS.

Where a standard database is a collection of tuples in $D^n$, a
weighted database (as defined in Proserpio et
al.~\cite{proserpio2014calibrating}) is a function from a tuple to its
weight ($D \rightarrow \mathbb{R}$). In this setting, counting the
number of tuples with a particular property is analogous to summing
the weights of all such tuples. Counting queries can therefore be
performed using summations.

In fact, summing weights in a weighted dataset produces exactly the
same result as the corresponding counting query on the original
dataset, when the query does not contain joins. When the query does
contain joins, WPINQ scales the weight of each row of the join's
output to maintain a sensitivity of 1. Proserpio et
al.~\cite{proserpio2014calibrating} define the weight of each row in a
join as follows, {\revisioncolor where $A_k$ is the weights of rows of relation $A$ with join key $k$}:

\begin{equation}
  A {\bowtie} B = \sum_k \frac{A_k \times B_k^T}{||A_k|| + ||B_k||}
  \label{eqn:wpinq_join}
\end{equation}
\smallskip

Since the scaled weights ensure a sensitivity of 1, Laplace noise
scaled to $1/\epsilon$ is sufficient to enforce differential
privacy. WPINQ adds noise with this scale to the results of the
weighted query.

In our \system implementation of WPINQ, we use the DBMS to track the weights associated with each column in computed relations. We can accomplish this by modifying the analyst's query to add weight a
column to each relation. Consider the transformation for a simple counting query, in which we initialize each weight to 1:

\begin{lstlisting}
SELECT COUNT(*) FROM trips
\end{lstlisting}
\mydownarrow
\begin{lstlisting}
SELECT SUM(weight)
FROM (SELECT *, 1 AS weight FROM trips)
\end{lstlisting}


\noindent This transformation adds a weight of 1 to each row in the
 table, and changes the \lstinline|COUNT|
aggregation function into a \lstinline|SUM| of the rows' weights.  The
correctness of this transformation is easy to see: as required by
WPINQ~\cite{proserpio2014calibrating}, the transformed query adds a
weight to each row, and uses \lstinline|SUM| in place of
\lstinline|COUNT|.

We can accomplish the second task (scaling weights for joins) by first
calculating the norms $||A_K||$ and $||B_k||$ for each key $k$, then
the new weights for each row using $A_k \times B_k^T$. For a join
between the \lstinline|trips| and \lstinline|drivers| tables, for
example, we can compute the norms for each key:
\begin{lstlisting}
WITH tnorms AS (SELECT driver_id, 
                       SUM(weight) AS norm
                FROM trips
                GROUP BY driver_id),
     dnorms AS (SELECT id, SUM(weight) AS norm
                FROM drivers
                GROUP BY id)
\end{lstlisting}

\noindent Then, we join the norms relations with the original results
and scale the weight for each row:

\begin{lstlisting}
SELECT ..., 
  (t.weight*d.weight)/(tn.norm+dn.norm) AS weight
FROM trips t, drivers d, tnorm tn, dnorm dn
WHERE t.driver_id = d.id
  AND t.driver_id = tn.driver_id
  AND d.id = dn.id
\end{lstlisting}

The correctness of this transformation follows from
equation~\eqref{eqn:wpinq_join}. The relation \lstinline|tnorms|
corresponds to $||A_k||$, and \lstinline|dnorms| to $||B_k||$.  For
each key, \lstinline|t.weight| corresponds to $A_k$, and
\lstinline|d.weight| to $B_k$.

Finally, we can accomplish the third task (adding Laplace noise scaled
to $1/\epsilon$) as a post-processing task. Our complete \system implementation defines a recursive rewriter that replaces table references with subqueries that initialize weights to 1, and joins with subqueries that update the weights as above. By modifying the analyst's query to track and update weights using the DBMS, our \system implementation enables WPINQ to scale to large datasets. We evaluate its performance in Section~\ref{sec:perf-overh-intr}.

\subsection{The Matrix Mechanism}
\label{sec:matrix-mechanism}

The matrix mechanism~\cite{li2015matrix} is another general approach for answering a set of counting queries. The insight behind the matrix mechanism is that the optimal way of answering a workload of counting queries might involve first answering a \emph{different} set of queries, then inferring the answers to the workload queries based on these answers. The matrix mechanism is defined in terms of three matrices: the workload queries, represented as a matrix; the \emph{strategy matrix}, which specifies the queries to submit to the database, and a matrix containing the answers to the queries in the strategy matrix. Given these three matrices, the method for answering the workload queries can be specified as a matrix multiplication.

We present an implementation of the matrix mechanism in \system in Figure~\ref{fig:matrixmech}. Its inputs are matrices representing the workload and strategy, and a list of SQL queries corresponding to the strategy queries. We use the Laplace mechanism to answer the strategy queries, then transform the results into a matrix representation and perform the matrix multiplication specified by Li et al.~\cite{li2015matrix} to obtain the workload results. The \lstinline[language=scala]|mpInverse| method on matrices implements the Moore–Penrose pseudoinverse.

\begin{figure}
\begin{lstlisting}[language=scala]
def matrixMech(workload: Matrix[Int], 
               strategyMat: Matrix[Int],
               strategyQs: List[String],
               epsilon: Double) = {
  // answer the queries in the cell list
  val answers = strategyQs.map { q =>
    laplaceMechClip(q, epsilon / strategyQs.length()) }

  workload times (strategy.mpInverse()
                    times colMatrix(answers))
}
\end{lstlisting}
  \caption{The Matrix Mechanism in \system}
  \label{fig:matrixmech}
\end{figure}

The challenge of determining an optimal set of strategy queries remains; Li et al.~\cite{li2015matrix} consider this an orthogonal problem, and provide some heuristics for developing good strategies. Each of these heuristics can be implemented as Scala functions to generate sets of strategy queries for our implementation of the matrix mechanism.






\subsection{Multiplicative Weights (MWEM)}
\label{sec:mult-weights-mwem}

The MWEM algorithm~\cite{hardt2012simple} is an iterative algorithm for answering a workload of counting queries with differential privacy. It provides a general algorithmic framework for iteratively improving a differentially private synthetic representation of the underlying data, until the synthetic representation is able to answer the queries in the workload with high accuracy.

Here, we develop an implementation of MWEM for 1-dimensional range queries over a single database table, based on a histogram representation of the data in the table. {\revisioncolor The full implementation appears in Figure~\ref{fig:mwem}.} To use the mechanism, the analyst provides a list of range queries (the workload), plus a list of ``bin edges'' which partition the domain of the table into histogram bins.

At each iteration of the algorithm, we perform two steps: (1) using the exponential mechanism, select a query from the workload which the synthetic representation \emph{cannot} answer with high accuracy; (2) using the Laplace mechanism, obtain a differentially private answer to this query, and use the \emph{multiplicative weights update rule} to update the synthetic representation. The mechanism returns the final synthetic representation.

To simplify the implementation we present here, we require the analyst to specify queries in terms of an upper and lower bound on the desired range. A query of the form $(l, u)$ on column \lstinline|c| of table \lstinline|T| is equivalent to the SQL query \lstinline|SELECT COUNT(*) FROM T WHERE l <= c AND c <= u|.



\begin{figure}
\begin{lstlisting}[language=scala]
def mwem(queries: List[(Double, Double)], 
         bins: List[Double], 
         numIters: Int, 
         epsilon: Double): List[(Double, Int)] = {
  // answer range query using synthetic representation
  def rangeQuerySyn(synRep: List[(Double, Int)], 
                    lower: Double, upper: Double) = {
    var count = 0
    for (i <- 0 to synRep.length()) {
      if (i <= lower && i < upper)
        count = count + synRep(i)
    }

    count
  }

  // answer range query using actual data
  def rangeQuery(lower: Double, upper: Double) = 
    Select Count(*) From T  
      Where C <= lower And C < upper

  // update rule for MWEM
  def mwemUpdate(lower: Double, upper: Double, 
                 synRep: MutableList[(Double, Int)], 
                 epsilon: Double) = {
    val realAnswer = DB.execute(rangeQuery(lower, upper))
                       + Util.Laplace(1 / epsilon)
    val synAnswer = rangeQuerySyn(synRep, lower, upper)

    val total = synRep.map(_._2).sum()
    for (i <- 0 to synRep.length()) {
      if (i <= lower && i < upper)
        synRep(i) = synRep(i) * 
         exp((realAnswer - synAnswer) / (2 * total))
    }
  }

  // initialize all counts to 100
  val synRep = bins.map(l => (l, 100))

  // split the privacy budget
  val epsilon_i = epsilon / numIters

  for (i <- 0 to numIters) {
    // pick the "worst" query in terms of accuracy
    val bQs = queries.map { case (l, u) =>
      rangeQuery(l, u) - rangeQuerySyn(synRep, l, u) }
    val qIdx = noisyMax(bQs, epsilon_i / 2)

    // update synthetic rep using selected query
    val l, u = queries(qIdx)
    mwemUpdate(l, u, synRep, epsilon_i / 2)
  }
}
\end{lstlisting}

  \caption{The MWEM Algorithm in \system}
  \label{fig:mwem}
\end{figure}

\end{document}